\documentclass[a4paper]{jpconf}
\usepackage{graphicx}
\begin{document}
\title{Non-Riemannian geometry: towards new avenues for the physics of modified gravity}

\author{Gonzalo J. Olmo$^{1,2}$ and D. Rubiera-Garcia$^{3,2}$}
\address{$^1$ Departamento de F\'{i}sica Te\'{o}rica and IFIC, Centro Mixto Universidad de
Valencia - CSIC. Universidad de Valencia, Burjassot-46100, Valencia, Spain}
\address{$^2$ Departamento de F\'isica, Universidade Federal da
Para\'\i ba, 58051-900 Jo\~ao Pessoa, Para\'\i ba, Brazil}
\address{$^3$ Center for Field Theory and Particle Physics and Department of Physics, Fudan University, 220 Handan Road, 200433 Shanghai, China}
\ead{gonzalo.olmo@csic.es, drubiera@fudan.edu.cn}

\begin{abstract}
Less explored than their metric (Riemannian) counterparts, metric-affine (or Palatini) theories bring an unexpected phenomenology for gravitational physics beyond General Relativity. Lessons of crystalline structures, where the presence of defects in their microstructure requires the use of non-Riemannian geometry for the proper description of their properties in the macroscopic continuum level, are discussed. In this analogy, concepts such as wormholes and geons play a fundamental role. Applications of the metric-affine formalism developed by the authors in the last three years are reviewed.
\end{abstract}

\section{Introduction and motivation}

Despite the tremendous experimental success of General Relativity (GR) \cite{Will}, it is widely assumed that at the Planck scale, $l_P \sqrt{ \equiv \hbar G/c^3} \simeq 10^{-35}m$, it will be superseded by a quantum theory of gravity. At this scale, dramatic changes are expected to arise. Among the proposals in the literature one finds, for example, those of non-commutative geometry, quantized geometry, or higher-dimensional scenarios (e.g. string theory). It is natural to wonder what kind of frameworks could be suitable for an effective description in the low-energy limit of such a theory. In this sense, since the number of physically accessible observables in this regime might be limited and  observations have a finite precision, at the end of the day one might only have access to some sector of the whole theory. Effective descriptions through classical models might be necessary and, indeed, they could serve as a guide for developing the underlying fundamental theories.

GR predicts the existence of black holes (BH), which entail singularities in the space-time \cite{singularities}, though the very definition of singularities remains still obscure. At this point one could either live with the existence of those singularities, postulate that quantum gravity somehow regularizes the metric, or look for self-consistent descriptions of the notion of body within GR. The latter path was taken by Wheeler \cite{Wheeler}, who formulated the concept of geon, namely, self-supported gravitational electromagnetic entities. Combined with the notion of multiply-connected topology, Misner and Wheeler \cite{MW} showed that through a wormhole a sourceless electromagnetic field can create mass and charge. Self-gravitating geons and non-Euclidean topologies are thus classical concepts with non-trivial consequences for quantum gravity. They indeed play a role within a larger picture: that of space-time foam \cite{stf}. If topology change can occur dynamically in Nature, then the smoothness of Minkowski space would disappear at Planckian scales, where not only geometry, but also topology, may fluctuate in all possible manners. Quantum fluctuations would lead then to creation and annihilation of wormholes. This raises a question: what kind of framework should be used to represent a scenario where a macroscopic, geometrical, continuum description arises from a discrete, microscopic structure?.

Crystalline structures and, in particular, Bravais crystals, provide valuable lessons to address the above question. The microscopic structure of crystals is represented by a discrete net-like arrangement of atoms, whose respective separation is given by a lattice constant. However, at the macroscopic level the crystal is naturally described in terms of differential geometry, resulting from a continuizing procedure in which the lattice spacing is taken to zero, while keeping the point mass unchanged. At each atom of the microscopic network one can define a triad of vectors (crystallographic directions); moving along these vectors we jump from atom to atom. Distances can be measured by step counting along crystallographic directions, $ds^2=g_{ij} dx^i dx^j$. A connection $\Gamma_{\mu\nu}^{\lambda}$ is used to transport vectors and define geodesics. For ideal and perfect crystals (the latter being just a deformation of the former) the macroscopic geometry can be described by a Riemannian structure, where the connection is compatible with the metric, $\nabla^{\Gamma} g_{\alpha\beta}=0$.

It turns out, however, that all known ordered structures contain defects. These are defined as a breakdown of the regular network pattern. They can be of different types, e.g., point-like (vacancies and intersticials), line-like (dislocations/disclinations) or two-dimensional (textures). Such defects are very important for the macroscopic description of materials, since they largely determine collective properties such as viscosity or plasticity. Here we shall deal only with point defects. Intersticials corresponds to atoms that have abandoned their equilibrium positions, while vacancies are the empty places left behind. At the location of the defects the step-counting procedure breaks down. The situation becomes non-metric, $\nabla^{\Gamma} g_{\alpha\beta} \neq 0$. This defected crystal can be also continuized to the macroscopic level, provided that, in addition, the density of defects is kept finite. The new description turns out to be given in terms of a metric-affine geometry \cite{Kroner}. To overcome the presence of defects, a new metric, $h_{\mu\nu}$, that ignores intersticials and fills vacancies, allows to step-count in the defected crystal, $\nabla^{\Gamma} h_{\alpha\beta}=0$. The deformation relating $g_{\mu\nu}$ and $h_{\mu\nu}$ depends on the kind and density of defects. Similarly, Cartan's torsion, $S_{\beta\gamma}^{\alpha}=\Gamma_{\beta\gamma}^{\alpha}-\Gamma_{\gamma\beta}^{\alpha}$, represents the continuum version of crystal dislocation. In summary, systems with defects require a non-Riemannian description.

\section{Main framework of metric-affine approach.}

Let us consider a gravitational action of the form
\begin{equation} \label{eq:action}
S=\frac{1}{2\kappa^2} \int d^4x \sqrt{-g} f(g_{\mu\nu},{R^\alpha}_{\beta\mu\nu}) + S_m(g_{\mu\nu},\Psi) \ ,
\end{equation}
where $\kappa^2 =8\pi G$ represents Newton's constant, $f$ is a given function of objects constructed with the Riemann tensor ${R^\alpha}_{\beta\mu\nu}$ and its contractions with the space-time metric $g_{\mu\nu}$ (where $g$ is its determinant), and $S_m$ is the matter action with $\Psi_m$ representing collectively the matter fields. Note that the connection $\Gamma_{\mu\nu}^{\lambda}$ does not couple to the matter sector. As we are working in the metric-affine (or Palatini) approach, the metric $g_{\mu\nu}$ and the affine connection $\Gamma_{\mu\nu}^{\lambda}$ are taken as independent objects and not constrained a priori. The dynamics of such theories is, in general, inequivalent to that of the standard metric (Riemannian) approach \cite{Borunda}.

Typically the function $f$ in (\ref{eq:action}) will be a combination of curvature invariants such as $R=g_{\mu\nu}R^{\mu\nu}$, $Q=R_{\mu\nu}R^{\mu\nu}$ and $K={R^\alpha}_{\beta\gamma\delta}{R_\alpha}^{\beta\gamma\delta}$ ($R_{\mu\nu}\equiv{R^\alpha}_{\mu\alpha\nu}$). In this section we focus on the family of theories $f(R,Q)$. To implement the Palatini approach we perform independent variation of the action (\ref{eq:action}) with respect to metric and connection, which yields \cite{olmo}

\begin{eqnarray}
f_R R_{\mu\nu}-\frac{f}{2}g_{\mu\nu}+2f_QR_{\mu\alpha}{R^\alpha}_\nu &=& \kappa^2 T_{\mu\nu}\label{eq:met-varX}\\
\nabla_{\beta}^\Gamma\left[\sqrt{-g}\left(f_R g^{\mu\nu}+2f_Q R^{\mu\nu}\right)\right]&=&0  \ ,
 \label{eq:con-varX}
\end{eqnarray}
where $f_R \equiv df/dR$, $T_{\mu\nu}$ is the energy-momentum tensor, and vanishing torsion, $\Gamma_{[\mu\nu]}^{\lambda}=0$, has been imposed \cite{torsion}. Tracing in (\ref{eq:met-varX}) with the metric $g^{\mu\nu}$ one obtains that $R=R({T_\mu}^{\nu})$ and $Q=Q({T_\mu}^{\nu})$ only depend on the matter sources. This allows us to introduce a new rank-two tensor, $h_{\mu\nu}$, satisfying $\nabla_{\beta}^\Gamma(\sqrt{-h} h^{\mu\nu})=0$. From Eq.(\ref{eq:con-varX}) the relation between $g_{\mu\nu}$ and $h_{\mu\nu}$ becomes

\begin{equation} \label{eq:h-g}
h^{\mu\nu}=\frac{g^{\mu\alpha}{\Sigma_{\alpha}}^\nu}{\sqrt{\det \hat{\Sigma}}} \ , \quad
h_{\mu\nu}=\left(\sqrt{\det \hat{\Sigma}}\right){{\Sigma^{-1}}_{\mu}}^{\alpha}g_{\alpha\nu} \ ,
\end{equation}
where the matrix ${\Sigma_\mu}^{\nu}=f_R \delta_{\mu}^{\nu} + 2f_Q {R_\mu}^{\nu}$ contains the relative deformation between them. Simple algebraic transformations allow us to rewrite (\ref{eq:met-varX}) as

\begin{equation} \label{eq:field}
{R_\mu}^{\nu}(h)=\frac{1}{\sqrt{\det \Sigma}} \left(\frac{f}{2} \delta_{\mu}^{\nu} + \kappa^2 {T_\mu}^{\nu} \right) \nonumber
\end{equation}
This constitutes a second-order system of differential field equations which in vacuum, ${T_\mu}^{\nu}=0$, reduce to those of GR plus a cosmological constant term (this term can be also generated through a non-linear electrodynamics model \cite{gorv}). Thus no extra propagating degrees of freedom arise. Note that since $g_{\mu\nu}$ is algebraically related to $h_{\mu\nu}$ via the matter sources only, $g_{\mu\nu}$ satisfies second-order equations as well. In the case of $f(R)$ theories ($f_Q=0$) only the trace of the energy-momentum tensor, $T=g_{\mu\nu}T^{\mu\nu}$, contributes to the modifications to the GR structure and, consequently, theories with non-vanishing trace, like the case of Born-Infeld non-linear electrodynamics  \cite{BI}, must be considered \cite{orfR}.

In summary, in Palatini $f(R,Q)$ gravity, the matter built up a Riemannian structure associated to $h_{\mu\nu}$. The structure for $g_{\mu\nu}$, however, is non-Riemannian, and the role of the deformation matrix ${\Sigma_\mu}^{\nu}$ is akin to that of point defects in crystalline structures. Once specific choices of gravity $f(R,Q)$ and matter $S_m$ are made, this framework provides a full solution to a given problem. Applications are now possible.

\section{Applications}

\subsection{A realization of Wheeler's geon}

Consider a matter source of the form

\begin{equation}
S_m=-\frac{1}{16\pi} \int d^4x \sqrt{-g} F_{\mu\nu}F^{\mu\nu}
\end{equation}
corresponding to a Maxwell field $F_{\mu\nu}=\partial_{\mu}A_{\nu} - \partial_{\nu}A_{\mu}$. The gravity Lagrangian is taken to be $f(R,Q)=R+l_P^2(aR^2+Q)$, where $a$ is a dimensionless constant. Let us consider static spherically symmetric solutions. Solving the field equations (\ref{eq:field}), these solutions can be cast, in ingoing Eddington-Finkelstein coordinates, as \cite{or12}

\begin{equation}\label{eq:ds2_EF}
ds^2=-A(x)dv^2+\frac{2}{\sigma_+}dvdx+r^2(x)d\Omega^2 \ ,
\end{equation}
with the definitions
\begin{eqnarray}\label{eq:A}
A(x)&=& \frac{1}{\sigma_+}\left[1-\frac{r_S}{ r  }\frac{(1+\delta_1 G(r))}{\sigma_-^{1/2}}\right] \\
\delta_1= \frac{1}{2r_S}\sqrt{\frac{r_q^3}{l_P}}\hspace{0.2cm};\hspace{0.2cm}
\sigma_\pm&=&1\pm \frac{r_c^4}{r^4(x)}\hspace{0.2cm};\hspace{0.2cm}
r^2(x)=\frac{x^2+\sqrt{x^4+4r_c^4}}{2} \label{eq:r(x)} \ ,
\end{eqnarray}
where $r_S=2M$ is Schwarzschild radius and $r_c=\sqrt{r_ql_P}$, with $r_q=2Gq^2$ a charge scale. A few Planck scale units away from the surface $x=0$ one has $r^2(x) \approx x^2$ and the geometry reduces to the standard Reissner-Nordstr\"om (RN) solution of GR. However, as the surface $x=0$ (or $r=r_c$) is approached, which replaces the point-like singularity of the RN solution, important non-perturbative modifications occur. For arbitrary values of the mass and charge curvature divergences arise on a sphere of radius $A=4\pi r_c^2(x)$ (though they are much milder than in the RN case), but they disappear when a certain charge-to-mass ratio is satisfied. The surface $x=0$ ($r=r_c$) represents the mouth of the wormhole, a tunnel to another region of space-time. The presence of such a wormhole allows to define an electric charge in terms of lines of electric flux flowing through $x=0$, though no point-like charges are present in the system. These facts allow to consistently interpret these solutions as explicit realizations of Wheeler's geon. Their existence may have important phenomenological consequences for the dark matter problem \cite{phem1}, for objects at particle accelerators \cite{phem2} and for the quantum foam picture \cite{phem3}.

\subsection{Dynamical formation}

Consider now an ingoing flux of particles carrying mass and charge, $T_{\mu\nu}^{flux}=\rho_{in} k_{\mu}k_{\nu}$, where $k_{\mu}$ is a null vector, $k_{\mu}k^{\mu}=0$, and $\rho_{in}$ is the flux density. Upon interaction with this flux, the static geonic solutions above get replaced by a generalization of the Bonnor-Vaidya solution of GR \cite{BV} of the form \cite{lor}

\begin{equation}
ds^2=-\left[\frac{1}{\sigma_+}\left(1-\frac{1+\delta_1 (v) G(z)}{\delta_2(v) z \sigma_{-}^{1/2}}\right)- \frac{2l_P^2 \kappa^2 \rho_{in}}{\sigma_{-}(1-\frac{2r_c^4}{r^4})}\right]dv^2 + \frac{2}{\sigma_+}dvdx+r^2(x,v) d\Omega^2 \ , \label{eq:ds2final}
\end{equation}
where the time-dependent $M(v)$ and $q(v)$ functions reflect the change in mass and charge due to the action of the ingoing flux. Assuming a charged perturbation in a compact interval $[v_i,v_f]$, the metric (\ref{eq:ds2final}) describes either the process of formation (if the initial state $v_i$ is Minkowski space) or the change in the geometry of the wormhole whose throat has an area $A=4\pi r_c^2(v_f)$. Apparently the first process implies a topology change, though this may not be the case if one assumes that the exact case $q=0$ is never physically realized and that only the limit $q \rightarrow 0$ is physically meaningful.

\subsection{Born-Infeld gravity and its extensions}

A different example of Palatini gravity is the proposal by Deser and Gibbons \cite{Deser}, given by the Lagrangian density

\begin{equation} \label{eq:actionBI}
S_{BI}=\frac{1}{\kappa^2\epsilon}\int d^4x \left[\sqrt{-|g_{\mu\nu}+\epsilon R_{\mu\nu}(\Gamma)|}-\lambda \sqrt{-|g_{\mu\nu}|}\right]
+S_m(g_{\mu\nu},\psi_m)
\end{equation}
dubbed, due to its resemblance with the non-linear electromagnetic theory, Born-Infeld (BI) gravity. When expanded in series of the (length-squared) parameter $\epsilon \ll 1$, the Lagrangian (\ref{eq:actionBI}) reduces, at first order, to the quadratic $f(R,Q)$ Lagrangian studied above. The representation of the field equations (\ref{eq:field}) of a $f(R,Q)$ theory is also valid for this Lagrangian formulated in the Palatini approach. In addition, for an electromagnetic field, all the subsequent orders in the above expansion exactly cancel out due to the symmetries of the energy-momentum tensor or the matter. As a consequence, BI gravity also supports (for $\epsilon<0$) geon-like solutions \cite{ors}.

The Lagrangian density (\ref{eq:actionBI}) admits an alternative representation as $L_{BI}=\sqrt{\det \hat{\Omega}}$, where the matrix $\hat{\Omega}$ is defined as ${\Omega^\alpha}_{\nu}=g^{\alpha\beta}q_{\beta \nu}$, where $q_{\mu\nu} \equiv g_{\mu\nu}+\epsilon R_{\mu\nu}(\Gamma)$. This representation suggests that BI gravity is just a particular case of a larger family of Lagrangians, given by $f(A)$, with $A=\det (\hat{\Omega})$. In particular, the family $f=A^{n/2}$ has been recently shown \cite{or-cosmo} to support bouncing cosmological solutions for different values of $n$ and different equations of state, which confirms the robustness of this kind of theories to avoid the Big Bang singularity as well.

\subsection{Higher-dimensional solutions}

One major motivation to extend metric-affine theories to extra dimensions is provided by potential applications within the AdS/CFT correspondence. The simplest case is that of $f(R)$ gravity in five space-time dimensions. The field equations (\ref{eq:field}), with $f_Q=0$, are valid also in the higher-dimensional case \cite{blor}. This means that the field equations of Palatini gravity are second-order in any dimension, without any need to constrain a priori the gravity Lagrangian or to consider ad hoc approaches such as those of quasi-topological gravities \cite{quasi-topological}. Note that in arbitrary space-time dimension, $n$, the relation between $h_{\mu\nu}$ and $g_{\mu\nu}$ becomes now

\begin{equation}
h_{\mu\nu}=f_R^{\frac{2}{n-2}} g_{\mu\nu} \hspace{0.1cm}; \hspace{0.1cm} h^{\mu\nu}=f_R^{\frac{2}{2-n}} g^{\mu\nu},
\end{equation}
so they are conformally related and depending on the number of space-time dimensions. In addition, $\det (\Sigma) =f_R^{\frac{4}{n-2}}$. For an electromagnetic field, solutions for a given $f(R)$ model can be obtained and expanded in series in the region of interest, with the result that, though curvature divergences appear, similar discussion as in the four-dimensional case brings that wormhole solutions may exist \cite{blor}. The analytic tractability of this framework, and the second-order character of the field equations, could therefore used to investigate concrete proposals within the AdS/CFT scenario. Further extensions to other theories of gravity such as $f(R,Q)$ or Born-Infeld gravity are also possible.

\subsection{Thick branes}

To reconciliate the apparent contradiction between the addition of extra dimensions with the fact that Nature is four-dimensional, scenarios where the extra dimension is very small (Kaluza-Klein models) or where it is infinitely large has been introduced. One example of the latter is the Randall-Sundrum scenarios, specifically the RS2 one \cite{RS2}, where one assumes that the four space-time dimensional world is embedded in an AdS5 warped geometry, with a single extra spatial dimension of infinite extension. Adding scalar fields to this scenario gives rises to thick branes, in which the warp factor depends on the dynamics of the scalar field, and can be used to control the splitting of the brane. We consider a standard scalar field, $L_S=\frac{1}{2} g^{\mu\nu} \partial_{\mu}\phi \partial_{\nu} \phi + 2V(\phi))$ and a space-time metric $ds^2=e^{2A(y)} g_{ab} dx^a dx^b + dy^2$, where $A(y)$ is the warp factor. A first-order formalism can be implemented to solve the corresponding Palatini field equations, through the introduction of a superpotential $W(\phi)$ such that $\theta= -\frac{1}{3}f_R^{1/3} W$, where we have defined $\theta \equiv A_y + \frac{1}{3} \frac{f_{R,y}}{f_R}$. The corresponding first-order equations are thus obtained as \cite{blror}

\begin{equation}
\phi_y=\frac{f_R^{4/3}}{\kappa^2} W_\phi \hspace{0.1cm}; \hspace{0.1cm} A_y= -\frac{1}{3}f_R^{1/3} \left(W+\frac{1}{\kappa^2}f_{R,\phi}W_\phi\right) \label{eq:Ay}\label{eq:Ay}\,.
\end{equation}
and the curvature reads $\frac{R}{20}=\frac{f_R^{2/3}}{3}\left(\frac{2}{5\kappa^2}f_R W_\phi^2-\frac{1}{3}W^2\right)$. Specifying a pair ($W(\phi), f_R(\phi)$) these equations allow to obtain the warp factor $A(y)$. It has been shown \cite{blror}, both perturbatively and with an exact solution, that a general property of these Palatini branes is that the warp factor tends to localize extra dimension, a feature not previously observed in this kind of models.

\section{Final remarks}

Metric-affine theories have an unexpected high-energy phenomenology and might contribute to shed light on the approaches to quantum gravity. The lessons from crystalline structures show that a macroscopic, continuum description in terms of differential geometry may arise from a discrete microscopic structure. Since the latter contain defects, this supports the view that perfect (Riemannian) structures might be unstable, and that a description in terms of metric-affine (non-Riemannian) structures is physically more sensible. Insights on how these lessons can be useful to unveil properties of the microscopic structure of space-time have been already obtained in \cite{lor14}, though many open questions remain and further research is needed.
%further research is needed to identify potential observational signatures resulting from the existence of macroscopic properties dictated by the presence of defects (wormholes) in the microstructure of space-time.
As a final lesson, we point out that determining whether the structure of space-time is Riemannian or not, is a question to be answered by experiment, not by tradition or convention.

\section*{Acknowledgments}

G.J.O. is supported by the Spanish grant FIS2011-29813-C02-02, the Consolider Program CPANPHY-1205388, the JAE-doc program of the Spanish Research Council (CSIC), and the i-LINK0780 grant of CSIC. D.R.-G. is supported by the NSFC (Chinese agency) grant No. 11305038, the Shanghai Municipal Education Commission grant for Innovative Programs No. 14ZZ001, the Thousand Young Talents Program, and Fudan University. The authors also acknowledge funding support of CNPq (Brazilian agency) project No. 301137/2014-5. The authors are grateful to Dionisio Bazeia, Eduardo Guendelman, Francisco S. N. Lobo, Laercio Losano, Jesus Martinez-Asencio, Roberto Menezes, Sergei Odintsov, Helios Sanchis-Alepuz and Mahary Vasihoun for their collaboration in several parts of this framework.

\end{document}